\documentclass[twocolumn,showpacs,amsmath,amssymb,superscriptaddress]{revtex4-1}

\usepackage{dcolumn}

\usepackage{color}
\usepackage{graphicx}
\usepackage{dcolumn}
\usepackage{bm}

\begin{document}


\title{Robust Regularity in $\gamma$-Soft Nuclei and its Microscopic Realization}

\author{K.~Nomura}
\affiliation{Department of physics, University of Tokyo, Hongo,
Bunkyo-ku, Tokyo 113-0033, Japan} 

\author{N.~Shimizu}
\affiliation{Center for Nuclear Study, University of Tokyo, Hongo,
Bunkyo-ku, Tokyo 113-0033, Japan} 

\author{D.~Vretenar}
\affiliation{Physics Department, Faculty of Science, University of
Zagreb, 10000 Zagreb, Croatia}

\author{T.~Nik$\check{\rm s}$i\'c}
\affiliation{Physics Department, Faculty of Science, University of
Zagreb, 10000 Zagreb, Croatia} 

\author{T.~Otsuka}
\affiliation{Department of physics, University of Tokyo, Hongo,
Bunkyo-ku, Tokyo 113-0033, Japan} 
\affiliation{Center for Nuclear Study, University of Tokyo, Hongo,
Bunkyo-ku, Tokyo 113-0033, Japan} 
\affiliation{National Superconducting Cyclotron Laboratory, 
Michigan State University, East Lansing, Michigan 48824-1321, USA}

\date{\today}

\begin{abstract}
$\gamma$-softness in atomic nuclei is investigated in 
the framework of energy density functionals. 
By mapping constrained microscopic energy surfaces for a set of
representative non-axial medium-heavy and heavy nuclei to a Hamiltonian of the
proton-neutron interacting boson model (IBM-2) containing up to three-body interactions, 
low-lying collective spectra and transition rates are calculated. 
Observables are analyzed that distinguish between the two limiting 
 geometrical pictures of non-axial nuclei: the rigid-triaxial rotor 
and the $\gamma$-unstable rotor.  It is shown that neither of these pictures 
is realized in actual nuclei, and that a microscopic description leads
 to results that are almost exactly in between the two geometrical
 limits. This finding points to the optimal choice of 
 the IBM Hamiltonian for $\gamma$-soft nuclei.  
\end{abstract}


\maketitle



Like many other quantum systems, atomic nuclei display a variety of 
geometrical shapes that
reflect deformations of the nuclear surface arising from 
collective motion of many nucleons \cite{BM}. 
Shapes of most 
non-spherical nuclei are characterized by axially-symmetric quadrupole 
deformations -- prolate or oblate ellipsoids. 
There are, however, many nuclei in 
which axial symmetry, i.e., the invariance under 
rotation around the symmetry axis of the intrinsic state, is broken. 
The precise description of axially asymmetric shapes and  
the resulting triaxial quantum many-body rotors 
remains open questions in nuclear physics and, 
since they are also being developed for 
other finite quantum systems like polyatomic molecules \cite{Herzberg1945II},
presents a topic of broad interest. 

Quadrupole shape deformations can be 
described in terms of the polar
deformation parameters $\beta$ and $\gamma$ \cite{BM}. 
The parameter $\beta$ is proportional to the intrinsic quadrupole moment,
and the angular variable $\gamma$ specifies the type of the shape. 
The limit $\gamma = 0$ corresponds to axial
prolate shapes, whereas the shape is oblate for $\gamma = \pi/3$. 
Triaxial shapes are associated with intermediate values $0 < \gamma < \pi/3$.  
The latter have been investigated extensively using theoretical approaches 
that are essentially based on the rigid-triaxial rotor model of Davydov and Filippov 
\cite{triaxial}, and the $\gamma$-unstable rotor model of Wilets and Jean \cite{gsoft}. 
The former assumes that the collective potential has a stable minimum at 
a particular value of $\gamma$, whereas in the 
latter the potential is independent of $\gamma$ and thus 
the corresponding collective wave functions are extended in the $\gamma$ direction.

However, presumably all known axially-asymmetric
nuclei exhibit features that are almost exactly in between these two
geometrical limits, characterized by the energy-level pattern of quasi-$\gamma$ band: 
relative locations of the odd-spin to the even-spin levels. 
As the two models originate from different physical pictures, the question
of whether axially-asymmetric nuclei are $\gamma$ rigid or unstable has attracted 
considerable theoretical interest \cite{BM,IBM,RS,Casten_Book}.  
The present Letter addresses this question from a microscopic
perspective, and identifies the appropriate Hamiltonian of the
interacting boson model (IBM) \cite{IBM} for $\gamma$-soft nuclei, 
consistent with the microscopic picture. 
We thereby provide a solution to the problem concerning the
energy-level pattern of the odd-spin states. 

\begin{figure}[ctb!]
\begin{center}
\includegraphics[width=8.3cm]{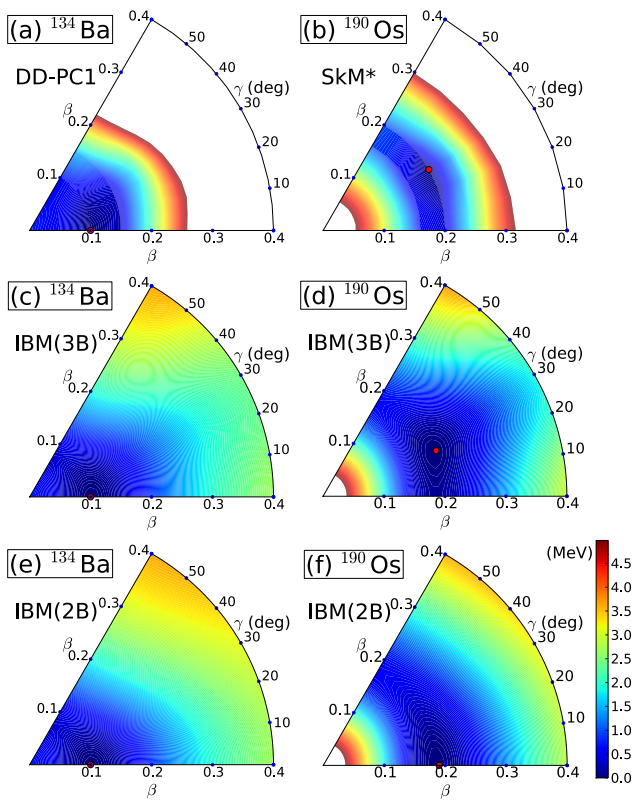}
\caption{(Color online) Self-consistent energy surfaces of 
$^{134}$Ba (a) and $^{190}$Os (b) up to 5 MeV in energy from the minima
 with the DD-PC1 and with the Skyrme SkM* functionals, respectively. 
The corresponding mapped energy surfaces of the IBM with (middle row)
 and without (lower row) the three-body term of Eq.~(\ref{eq:cubic}) are
 plotted. 
} 
\label{fig:pes}
\end{center}
\end{figure}

At present the most complete 
microscopic description of ground-state properties 
and collective excitations over the whole chart of nuclides is provided by the 
framework of energy density functionals (EDFs). 
Both non-relativistic \cite{BHR.03,EKR.11,Sk-VB,Go}, 
and relativistic \cite{VALR.05,RMF_review} EDFs have successfully been employed in numerous 
studies of shape phenomena and the resulting complex excitation
spectra and decay patterns 
\cite{nso,Bender2008GCMtriaxial,TRodriguez2010triaxial,Ni.07}. The starting point is usually 
a constrained self-consistent mean-field calculation of the energy surface with the mass 
quadrupole moments as constrained quantities \cite{RS}. This is illustrated in the first row of 
Fig.~\ref{fig:pes}, where we display the self-consistent quadrupole
energy surfaces of $^{134}$Ba (a) and $^{190}$Os (b) in the 
$\beta - \gamma$ plane. 
The constrained energy surface of $^{134}$Ba is calculated using the relativistic
Hartree-Bogoliubov model \cite{VALR.05} with the DD-PC1 \cite{DD-PC1} functional, 
and that of $^{190}$Os employing the Hartree-Fock plus BCS model \cite{ev8}
with the Skyrme functional SkM* \cite{SkM}. These functionals are representative 
of the two classes -- relativistic and non-relativistic EDFs, and will be used throughout 
this work to demonstrate that the principal conclusions do not depend on the particular 
choice of the EDF. One notices in Figs.~\ref{fig:pes}(a) and \ref{fig:pes}(b) that in both cases the
energy surface is very soft in $\gamma$, with $^{134}$Ba displaying a nearly $\gamma$-independent 
picture, whereas a more pronounced rigid triaxial shape is predicted for $^{190}$Os 
with the minimum at $\gamma\approx 30^{\circ}$.

To calculate excitation spectra and transition rates, it is necessary to go beyond the
mean-field solution and explicitly take into account collective correlations. 
Here we employ the IBM to analyze spectroscopic properties 
of $\gamma$-soft nuclei. 
The O(6) dynamical symmetry \cite{IBM} of IBM presents
a good approximation to a system with $\gamma$-independent collective potential.  
The geometrical picture of the  O(6) limit of the IBM emerges in the 
 coherent-state framework \cite{coherent}, being consistent with
the model of Wilets and Jean \cite{gsoft}. 
The coherent state represents the intrinsic wave function of the boson system, 
and O(6) states in the laboratory system can be generated by  angular momentum 
projection \cite{coherent}. 
The triaxial-rotor features of the IBM  were emphasized already  
in \cite{Casten1984triaxial,Castanos1984effgamma}, leading
to the ``equivalence'' ansatz of the $\gamma$-rigid and the O(6) 
descriptions of the low-lying spectra \cite{Otsuka1987triaxial}.  

The present study uses the proton-neutron IBM (IBM-2), which includes 
proton (neutron) monopole $s_{\pi}$ ($s_{\nu}$) and quadrupole $d_{\pi}$ ($d_{\nu}$) bosons, 
representing $J^{\pi}=0^{+}$ and $2^{+}$ collective pairs of valence protons 
(neutrons), respectively \cite{OAI}. 
The number $N_{\pi}$ ($N_{\nu}$) of proton (neutron) bosons 
equals the number of valence proton (neutron) pairs (particles or holes), 
with respect to the nearest proton (neutron) closed shell 
\cite{OAI}. 
The following IBM-2 Hamiltonian is employed: 
\begin{eqnarray}
 H_{\rm IBM}=\epsilon (n_{d\pi}+n_{d\nu})+\kappa Q_{\pi}\cdot Q_{\nu}+H_{\rm 3B}
\label{eq:IBMhamiltonian}
\end{eqnarray}
with the $d$-boson number operator 
$n_{d\rho}=d^{\dagger}_{\rho}\cdot\tilde d_{\rho}$ ($\rho=\pi,\nu$),  
and the quadrupole operator 
$Q_{\rho}=s^{\dagger}_{\rho}\tilde
d_{\rho}+d^{\dagger}_{\rho}s_{\rho}+\chi_{\rho}[d^{\dagger}_{\rho}\tilde
d_{\rho}]^{(2)}$. 
The third term $H_{\rm 3B}$ on the right-hand side of
Eq.~(\ref{eq:IBMhamiltonian}) denotes the three-body boson interaction:   
\begin{eqnarray}
H_{\rm 3B}=
\sum_{\rho\neq\rho^{\prime}}
\sum_{L}
\theta^{\rho}_{L}
[d^{\dagger}_{\rho}d^{\dagger}_{\rho}d^{\dagger}_{\rho^{\prime}}]^{(L)}
\cdot [\tilde d_{\rho^{\prime}}\tilde d_{\rho}\tilde d_{\rho}]^{(L)}. 
\label{eq:cubic}
\end{eqnarray}
The three-body term was introduced and analyzed in the
IBM-1 framework (without distinction between proton and neutron
bosons) \cite{Isacker1981coherent,Heyde1984cubic}, but is used here 
for the first time in the microscopic IBM-2 model.  
In the IBM-2 there could be three-body terms with combinations 
of proton and neutron $d$-boson operators different from the one used in Eq.~(\ref{eq:cubic}). 
However, since the proton-neutron quadrupole interaction dominates over
the proton-proton and neutron-neutron ones for medium-heavy and heavy
deformed nuclei, the term (\ref{eq:cubic}) represents the 
dominant contribution of three-body boson interactions. 
For each $\rho$ and $\rho^{\prime}$, there are five linearly independent combinations in  
Eq.~(\ref{eq:cubic}), determined by the value of $L=0,2,3,4,6$ \cite{Isacker1981coherent}. 
However, only the term with $L=3$ can give rise to a stable triaxial minimum at 
$\gamma\approx 30^{\circ}$ \cite{Heyde1984cubic}, because its expectation value 
in the classical limit is proportional to $\cos^{2}{3\gamma}$. We thus consider only the 
$L=3$ in Eq.~(\ref{eq:cubic}) and, in addition, assume $\theta^{\pi}_{3}=\theta^{\nu}_{3}\equiv \theta_{3}$. 

The parameters $\epsilon$, $\kappa$, $\chi_{\pi,\nu}$ and $\theta_{3}$ are 
adjusted following the procedure of
Ref.~\cite{nso}: the microscopic quadrupole  
energy surface, obtained from a mean-field calculation using a given EDF, 
is mapped onto the corresponding boson energy surface, i.e., expectation value of 
$H_{\rm IBM}$ in the coherent state (cf. \cite{nso,nsofull} for details). 
The deduced value of $\theta_{3} > 0 $ varies gradually with boson number: 
$|\theta_{3}/\kappa|\approx 1$ for $1\leqslant N_{\pi}+N_{\nu}\lesssim 5$
and $\approx 0.5$ for $5\lesssim N_{\pi}+N_{\nu}\leqslant 10$. 

Without three-body boson terms the energy expectation value either has a minimum at 
$\gamma=0^{\circ}$ (prolate shapes) or 60$^{\circ}$ (oblate shapes), or is independent 
of $\gamma$ in the O(6) limit. Triaxial minima are 
obtained only after the inclusion of the three-body interaction 
$H_{\rm 3B}$.  
This is nicely illustrated in Fig.~\ref{fig:pes}, where the mapped energy surfaces of the IBM are 
plotted in the middle row (for the full IBM Hamiltonian
Eq.~(\ref{eq:IBMhamiltonian}) that contains the three-body term), and
in the lower row (for the IBM Hamiltonian without the three-body term). 
For $^{190}$Os the Hartree-Fock plus BCS model with the Skyrme 
functional SkM* predicts a minimum 
at $\gamma\approx 30^{\circ}$, which can only be reproduced on the mapped 
surface corresponding to the expectation value of the full 
IBM Hamiltonian containing the three-body term (Fig.~\ref{fig:pes}(d)).  
The contribution of this term to the mapped energy surface 
is in general less important when the number of active bosons becomes relatively small. 
Thus for $^{134}$Ba nucleus in Fig.~\ref{fig:pes}(c) the minimum is still on prolate axis even
when the three-body term is included. 
The IBM Hamiltonian with up to two-body terms yields an energy surface that is soft in
the $\gamma$ degree of freedom (cf. Fig.~\ref{fig:pes}(f)), 
but the minimum is on the $\gamma=0^{\circ}$ axis. 
We note that while the angular variables $\gamma$ of the boson energy surface 
and the constrained microscopic energy surface 
are identical to each other, the axial deformation 
parameters $\beta$ are related by a constant of proportionality determined by equating 
the corresponding intrinsic quadrupole moments \cite{nso}. The geometrical variable 
$\beta$ is obtained by multiplying the boson axial deformation by factors 
$\approx 0.15$ and 0.2 for $^{134}$Ba and $^{190}$Os, respectively.

A distinction between $\gamma$-unstable and rigid-triaxial nuclei arises 
when considering the ratio of excitation energies \cite{Casten_Book}: 
 $S(J,J-1,J-2)\equiv [\{E(J)-E(J-1)\}-\{E(J-1)-E(J-2)\}]/E(2^{+}_{1})$ 
 for the quasi-$\gamma$ ($K^{\pi}=2^{+}$) band 
$J^{\pi}=2^{+}_{\gamma}$, $3^{+}_{\gamma}$, $4^{+}_{\gamma}$ \ldots, etc.  
The excitation energies $E(J)$ are obtained by diagonalization of the
Hamiltonian $H_{\rm IBM}$, and the quadrupole operators $Q_{\rho}$  are used 
in the calculation of E2 transition rates, with identical 
proton and neutron boson effective charges. 

\begin{figure}[ctb!]
\begin{center}
\includegraphics[width=8.3cm]{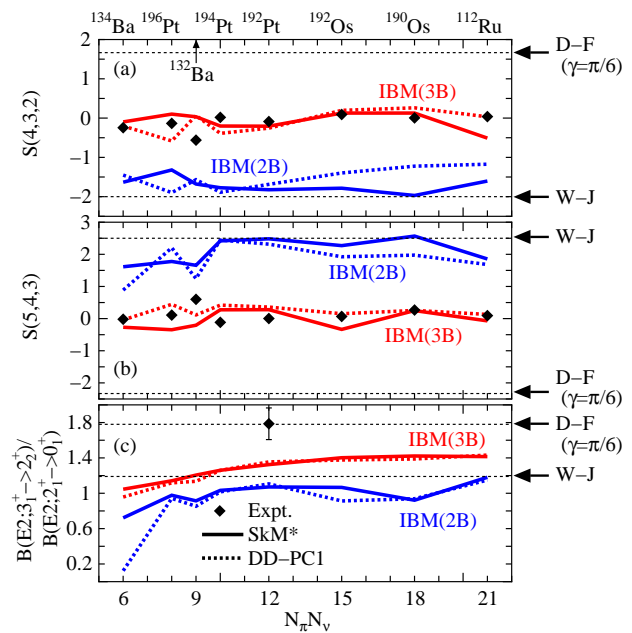}
\caption{(Color online) The energy ratios (a) $S(4,3,2)$ and (b) $S(5,4,3)$, 
and (c) the 
$B({\rm E2};3^{+}_{1}\rightarrow 2^{+}_{2})/B({\rm E2};2^{+}_{1}\rightarrow 0^{+}_{1})$ 
ratio, as functions of the product $N_{\pi}N_{\nu}$, 
for a characteristic set of non-axial medium and heavy nuclei. 
IBM(3B) and IBM(2B) denote results obtained with the 
IBM Hamiltonians with up to three- and two-body terms, respectively. The 
Skyrme SkM* and relativistic DD-PC1 functionals are used. 
Data are from Refs. \cite{data,192}, and D-F and W-J denote the limits
 of the rigid-triaxial and the $\gamma$-unstable (or O(6)) models, respectively. 
}
\label{fig:ratio}
\end{center}
\end{figure}

For a characteristic set of non-axial medium-heavy and heavy nuclei,  
in Fig.~\ref{fig:ratio} we plot the energy ratios $S(4,3,2)$ (a) and
$S(5,4,3)$ (b), as functions of the product of 
proton and neutron boson numbers: $N_{\pi}N_{\nu}$. 
The latter quantity reflects the amount of valence proton-neutron 
correlations, and hence the increase of $N_{\pi}N_{\nu}$ corresponds to
an enhancement of collectivity \cite{Casten_Book}. 
In this work we consider non-axial nuclei in the mass regions 
$A\sim 110$, 130 and 190, whose spectra display signatures 
of $\gamma$-softness. The set of nuclei shown in Fig.~\ref{fig:ratio} 
has been selected so that the corresponding values of $N_{\pi}N_{\nu}$ 
evenly span the widest possible range. 
The IBM excitation spectra have been calculated 
starting from self-consistent mean-field energy surfaces that correspond to 
the two functionals, Skyrme SkM* and the relativistic DD-PC1. 
The two energy ratios, calculated with and without the three-body 
term of Eq.~(\ref{eq:cubic}) in the IBM Hamiltonian, are plotted in
comparison to data \cite{data}, and the predictions of the rigid-triaxial rotor model of Davydov and Filippov 
\cite{triaxial} and the $\gamma$-unstable rotor model of Wilets and Jean \cite{gsoft}. 
One notices that for all considered nuclei data can only be reproduced with the 
IBM Hamiltonian that includes the three-body term Eq.~(\ref{eq:cubic}). Both the empirical 
and calculated ratios fall almost exactly in between the limits of the 
$\gamma$-unstable rotor and the rigid-triaxial rotor models: 
the Wilets-Jean limit is -2.00 and the
Davydov-Filippov limit is 1.67 for $S(4,3,2)$;  
the Wilets-Jean model predicts 2.50, and Davydov-Filippov -2.33 for $S(5,4,3)$. 
The IBM Hamiltonian with up to two-body terms cannot reproduce the 
empirical values and, in both cases, yields energy ratios that are  close to 
the predictions of the $\gamma$-unstable rotor model. 

While the energy ratios are largely independent of the product of boson numbers, 
the $B$(E2) systematics reflects the evolution of collectivity. For instance,  
the ratio 
$B({\rm E2};3^{+}_{1}\rightarrow 2^{+}_{2})/B({\rm E2};2^{+}_{1}\rightarrow 0^{+}_{1})$, 
plotted in Fig.~\ref{fig:ratio}(c), gradually increases with 
$N_{\pi}N_{\nu}$. 
For nuclei with typically low $N_{\pi}N_{\nu}$ ($\leqslant 10$), like
$^{132,134}$Ba and $^{194,196}$Pt, the $\gamma$ value on average is
close to $0^{\circ}$ or $60^{\circ}$. 
In this case the ratio 
$B({\rm E2};3^{+}_{1}\rightarrow 2^{+}_{2})/B({\rm E2};2^{+}_{1}\rightarrow 0^{+}_{1})$  
is closer to the Wilets-Jean limit (O(6) in the IBM representation) of 1.19. 
As the collectivity evolves with $N_{\pi}N_{\nu}\geqslant 12$, this 
$B$(E2) ratio, calculated with the full IBM Hamiltonian that includes 
the three-body term, saturates between the $\gamma$-rigid limit of 1.78 and 
the $\gamma$-unstable limit of 1.19, in agreement with behavior of the 
energy ratios $S(J,J-1,J-2)$. 
The $B$(E2) ratio calculated with the IBM Hamiltonian with up to
two-body terms remains close to the O(6) limit even for large values of 
$N_{\pi}N_{\nu}$.  

Although the ratios shown in Fig.~\ref{fig:ratio} are calculated using
two completely different microscopic density functionals, it appears
that the basic features of this analysis are not sensitive to the
particular choice of the underlying EDF.  

In the IBM picture, the number of proton (neutron) bosons equals half 
the number of the corresponding valence particles or holes \cite{OAI}. 
Among the nuclei discussed in this Letter, 
those with relatively large $N_{\pi}N_{\nu}$ ($\geqslant 12$), in many 
of which both $N_{\pi}$ and $N_{\nu}$ correspond to hole configurations, are more 
likely to exhibit pronounced $\gamma$ rigidity, compared to 
systems with low $N_{\pi}N_{\nu}$ ($\leqslant 10$). 
In most of the latter cases $N_{\pi}$ ($N_{\nu}$) corresponds 
to particle (hole) configuration, or vice versa.

\begin{figure}[ctb!]
\begin{center}
\includegraphics[width=8.3cm]{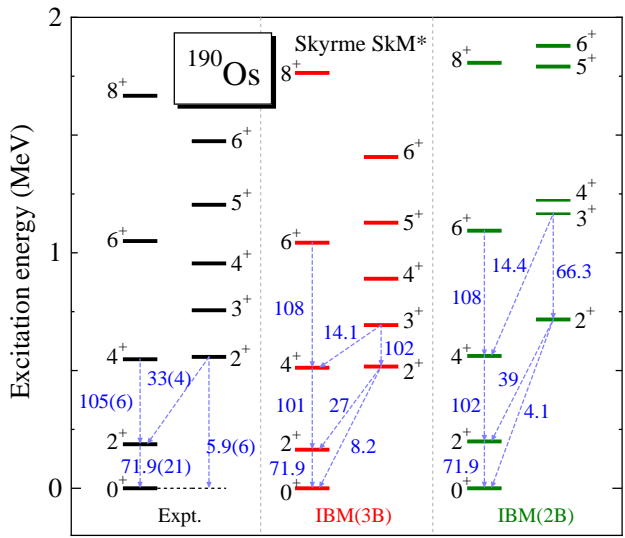} 
\caption{(Color online) Low-lying spectra and  $B$(E2)
 values (in Weisskopf units) of $^{190}$Os. The bands calculated with 
 IBM Hamiltonians with (IBM(3B)) and without (IBM(2B)) the 
three-body term of Eq.~(\ref{eq:cubic}) 
are compared to experimental data \cite{190}. The Skyrme SkM* EDF is used, 
and the boson effective 
charge is adjusted to reproduce the experimental value 
of $B({\rm E2};2^{+}_{1}\rightarrow 0^{+}_{1})$. }
\label{fig:spectra}
\end{center}
\end{figure}

The discussion so far has focused on the systematics of energy ratios and transition rates. 
The model, however, provides an equally accurate and complete description of low-energy 
excitation spectra in individual nuclei. 
This is highlighted by the level scheme of $^{190}$Os in
Fig.~\ref{fig:spectra}. 
Again we compare results obtained with IBM Hamiltonians containing up to two- and
three-body terms to available data \cite{data,190}. 
The full IBM Hamiltonian $H_{\rm IBM}$ reproduces 
both the excitation energies and transition rates for the ground-state band and 
the band built on the state $2^{+}_{2}$ (quasi-$\gamma$ band). 
We notice the marked effect of the three-body term on the quasi-$\gamma$
band: all states are lowered in energy but, in particular, the
pronounced lowering of the odd-spin states, 
e.g., $3^{+}_{1}$ and $5^{+}_{1}$ 
by 473 keV and 663 keV, respectively, breaks the quasi-degeneracy of the doublets 
($3^{+}_{1}$,$4^{+}_{2}$), ($5^{+}_{1}$,$6^{+}_{2}$), etc \cite{3+}. 
These doublets ($\tau$-multiplets) are 
characteristic of the $\gamma$-unstable O(6) limit of IBM \cite{IBM}.  
We emphasize that there are no additional adjustable parameters  
in the calculation of excitation energies, that is, 
the parameters are completely determined by the choice of the microscopic functional and 
the mapping procedure. Results of similar level of agreement with experiment are also 
obtained in the calculation of spectra of other nuclei considered in this study. 

In conclusion, we have investigated the emergence of $\gamma$ softness in atomic nuclei 
starting from the microscopic framework of energy density functionals. 
For a wide range of relevant 
nuclei certain observables allow us, 
in comparison to microscopic calculations, to differentiate 
two limiting geometrical pictures: the rigid-triaxial and the $\gamma$-unstable rotors. 
The present analysis clearly demonstrates that neither of these pictures is 
realized in actual nuclei. 
Typical non-axial medium-heavy and heavy 
nuclei lie almost exactly in the middle between the two geometrical 
limits, as a robust regularity. 
In the IBM framework the regularity arises naturally only when a 
three-body boson interaction is included. 
This result points to the origin of the three-body boson interaction,
suggesting the optimal IBM description of $\gamma$-soft 
nuclei. 
The principal results presented in this Letter do not depend on details of 
the EDF, and suggest us a comprehensive picture of triaxial shapes of
atomic nuclei in a fully microscopic way, including a solution
to the longstanding problem of the energy-level pattern of odd-spin states. 
%
%

This work has been supported in part by grants-in-aid for Scientific
Research (A) 20244022 and No.~217368, and by MZOS - project 1191005-1010.
K.N. and D.V. acknowledge support by the JSPS. 
T.N. acknowledges support by the Croatian Science Foundation.


\begin{thebibliography}{99}
\bibitem{3+}
The lowest 3$^+$ level is predicted too high in many IBM 
calculations for $\gamma$-soft nuclei, and the empirical lowering was
often ascribed to a possible coupling to two-quasiparticle states. 
The present work suggests that such a 
mechanism could play only a minor role.
\end{thebibliography}
\end{document}